\documentclass[twocolumn]{emulateapj}

\slugcomment{}

\shorttitle{The Dependence of SSFR and Its Dispersion on Galaxy Stellar Mass}
\shortauthors{Guo et al.}

\begin{document}

\title{The Star-Formation Main Sequence: The Dependence of Specific Star Formation Rate and Its Dispersion on Galaxy Stellar Mass}


\author{Kexin GUO\altaffilmark{1,2},Xian~Zhong ZHENG\altaffilmark{1},Tao WANG\altaffilmark{3,4} and Hai FU\altaffilmark{5}}

\altaffiltext{1}{Purple Mountain Observatory, Chinese Academy of Sciences, Nanjing 210008, China, kxguo@pmo.ac.cn, xzzheng@pmo.ac.cn}
\altaffiltext{2}{University of Chinese Academy of Sciences, Beijing 100049, China}
\altaffiltext{3}{Laboratoire AIM-Paris-Saclay, CEA/DSM/Irfu, Saclay, pt courrier 131, 91191 Gif-sur-Yvette, France}
\altaffiltext{4}{School of Astronomy and Space Science, Nanjing University, Nanjing 210093, China}
\altaffiltext{5}{Department of Physics and Astronomy, University of Iowa, Van Allen Hall, Iowa City, IA 52242, USA}

\begin{abstract}
The dispersion of the star-formation main sequence (SFMS) reflects the diversity of star formation histories and variation in star formation rates (SFRs) in star-forming galaxies (SFGs) with similar stellar masses ($M^\ast$) .
We examine the dispersion of local SFMS using a complete sample of Sloan Digital Sky Survey galaxies at  0.01$<z<$0.03 with $\log(M^\ast/M_\odot)>$8.8.
The SFRs are estimated from H$\alpha$ in combination with 22\,$\mu m$ observation from WISE.
We measure the dispersion of specific SFR (SSFR) as a function of $M^*$.
We confirm that the dispersion increases with $M^*$ from 0.37$\pm0.01$\,dex at $\log(M^\ast/M_\odot)<$9.6 to 0.51$\pm0.02$\,dex at $\log(M^\ast/M_\odot)>$10.2.
Despite star formation is mostly associated with disks, the dispersion of disk SSFR still increases with $M^*$.
We conclude that the presence of bulges/bars is likely responsible for the large dispersion of SSFR in massive SFGs while low-mass SFGs are mostly disk-dominated and thus with small dispersion.
Our results suggest that star formation on galactic scales is dramatically affected by central dense structures through both enhancing and/or quenching processes; while lower-mass SFGs tend to have less bursty star formation histories.
However, the dispersion of SSFR becomes significantly smaller and remains constant when only disk-dominated SFGs are counted. This finding implies that the impact of stochastic stellar feedback on star formation is likely to follow the same pattern in all disk galaxies,  showing no correlation with halo potential.
\end{abstract}

\keywords{galaxies: evolution --- galaxies: starburst --- infrared: galaxies}

\section{Introduction} \label{sec:intro}

Star-forming galaxies (SFGs) exhibit a tight correlation between stellar mass ($M^\ast$) and star formation rate (SFR), i.e. the so-called star formation main sequence (SFMS), from early cosmic epochs to the present day \citep{Noeske07, Elbaz07, Wuyts11, Karim11, Whitaker12}. This fundamental relationship has been widely used to test models of galaxy formation and evolution \citep[e.g.,][]{Peng10,Behroozi13}. While much progress has been made in understanding selection effects and uncertainties in SFR and $M^*$ measurements \citep{KE12, RP15}, the measurements of SFMS from different works based on multi-wavelength deep surveys surprisingly reached a good agreement when calibrated to the same standards \citep{Speagle14}.  Focus has now turned to characterizing the slope, dispersion and normalization of SFMS in detail, aimed at resolving the contributions of different physical processes regulating star formation on galaxy scales in a statistical sense.

The dispersion of SFMS at fixed $M^*$ measures the variation in the level of star formation among similarly massive galaxies. Measurement errors turn out to contribute little to the variation \citep{Salmi12} and the dispersion in specific SFR (SSFR=$SFR/M^\ast$) thus reflects the diversity in star formation histories (SFHs) and variability in SFRs on short time scales \citep{Dutton10,Sparre15}.

Multiple processes shape the SFH of a galaxy, including
gas accretion, minor mergers, stellar feedback and quasar/radio-mode feedback \citep[see][and reference therein]{Moustakas13}. Relative roles of these processes are usually dependent on halo mass and cosmic epoch \citep{Oser10,Behroozi13}. In particular,  stellar feedback is simulated to impact on further star formation, leading to a larger fluctuation in SFRs on time scales of $\sim10^{7-8}\rm\,yr$ in lower-mass galaxies \citep{Hopkins14}. However, this prediction is inconsistent with current observations. The dispersion of SSFR  is reported to be constant ($\sim$0.3\,dex) over a wide $M^*$ range for $z\sim 2$ SFGs \citep{Rodighiero11,Schreiber15},
but tends to be larger for more massive SFGs in lower-redshift universe \citep[][see also \citealt{Ilbert14}]{Guo13}.
SFR measurement for individual SFGs is critical in quantitatively studying the dependence of SSFR dispersion on $M^*$. More efforts are demanded to figure out what are responsible for the disagreement between theoretical and observational sides.

In this letter, we examine the dependence of SSFR dispersion on $M^*$ using a complete sample of local SFGs selected from the Sloan Digital Sky Survey (SDSS).  In addition, the presence of non-star-forming bulges in SFGs may lower SSFR and increase SSFR dispersion.
We take this effect into account in our examination. 
A Chabrier \citeyearpar{Chabrier03} Initial Mass Function (IMF) is used throughout this work.

\section{Sample and data} \label{sec:data}

We select a local sample of galaxies from SDSS data release 7 \citep{Abazajian09}, which covers $\sim8000\, \rm deg^2$ of the sky with spectroscopic observations nearly complete to $r<17.77$.
 It has become clear that $\rm H\alpha$+22\,$\micron$ is a robust SFR indicator with minimal scatter \citep[$\sim$ 0.14\,dex;][]{Kennicutt09,Hao11,Lee13} and the spread of SFRs is thus least influenced by the measurement errors. We therefore use the infrared (IR) 22\,$\micron$ observations from the Wide-field Infrared Survey Explorer (WISE) all-sky survey, together with $\rm H\alpha$ from SDSS to measure SFRs on short time scales ($\sim 10^7$\,yr), compared to the indicator of ultraviolet (UV)+IR continuum luminosity that measures SFRs over scales of $\sim 10^{8}$\,yr \citep{KE12}.
The SDSS catalog is cross-correlated with the WISE all-sky survey catalog using a matching radius of $3\arcsec$,  equal to half of the WISE beam size at 3.4$\,\micron$. Only the closest counterpart is chosen when multiple WISE sources are found within the radius. The WISE all-sky survey reaches a 5$\sigma$ sensitivity of 0.08, 0.11, 1 and 6\,mJy at 3.4, 4.5, 12 and 22\,$\micron$, respectively \citep{Yan13}.
We take 22\,$\micron$ fluxes above the 3\,$\sigma$ level ($\sim$3.6\,mJy) to estimate IR luminosity. This flux limit corresponds to an IR-based SFR of 0.1\,$M_\odot$\,yr$^{-1}$ at redshift $z=0.03$ when the IR spectral energy distribution (SED) of \citet{CE01} is applied, according to the average SFR of 10$^{9.5}\,M_\odot$ galaxies on local SFMS by \citet{Whitaker12}.
Our sample is limited to 0.01$<z<$0.03 in order to collect a sufficiently large number of galaxies with 22\,$\micron$ detection.  The lower limit $z=0.01$ is set to avoid serious fiber aperture effect \citep{Hopkins03,Lee13}. We adopt the photometric magnitudes in $u, g$ and $r$  from the SDSS pipeline \citep{Stoughton02}, and optical emission line fluxes, rest-frame colors, $M^*$ from the MPA-JHU value-added galaxy catalogs (VAGCs; \citealp{Tremonti04}), and S{\'{e}}rsic indices from the NYU value-added catalog \citep{Blanton05}. The stellar mass in VAGCs are derived from optical broadband SEDs. The SDSS limit of $r<17.77$ for the main galaxy sample enables a complete selection for galaxies with $\log(M^\ast/M_\odot) \geq$8.8 at $z<0.03$. We convert IMF from Kroupa to Chabrier by dividing $M^*$ a factor of 1.06.

After removing duplicated objects and spectrally classified AGNs, we obtain 21\,307 galaxies with $\log(M^\ast/M_\odot) \geq$8.8 in $0.01<z< 0.03$ as our sample.  Following \citet{Brinchmann04}, we correct fiber aperture effect for observed H$\alpha$ fluxes. In practice, we use the relationships between H$\alpha$ equivalent width and $u-g-r$ fiber colors to estimate H$\alpha$ flux
outside the fiber and obtain the total H$\alpha$ flux of a galaxy.
We test our aperture correction using a sample of SDSS galaxies at $0.05<z<0.1$, for which fiber aperture effect is ignorable, finding an uncertainty of 0.2\,dex for the total $\rm H\alpha$ flux to be independent of $M^*$.
We calculate SFR following \citet{Lee13} as below:
\begin{equation}\label{eq:1}
  {SFR=9.12\times10^{-9}\,(L_{\rm H\alpha}+0.034\,L_{22})^{1.06}},
\end{equation}
where $L_{22}$ is 22\,$\micron$ monolithic luminosity. A factor of 1.7 is used to convert Salpeter into Chabrier IMF.

For 22\,$\micron$-undetected galaxies, either quiescent ones or IR-faint SFGs mostly with $\log(M^\ast/M_\odot)<\sim$9.5, SFR is estimated from  H$\alpha$ flux. The observed H$\alpha$ flux is corrected for extinction, which is determined using Balmer decrement with intrinsic H$\alpha$/H$\beta$=2.86 and the extinction law of \citet{CCM89} \citep[with $R_{\rm V}=3.1$ and coefficients updated by][]{Od94}.
Then SFR is calculated using
\begin{equation}\label{eq:2}
{SFR=7.9\times10^{-42}\,L_{\rm H\alpha,corr}},
\end{equation}
where $L_{\rm H\alpha,corr}$ is extinction-corrected H$\alpha$ luminosity.
We note that  22\,$\micron$-undetected galaxies are marginally affected by extinction and the estimate of SFR from extinction-corrected H$\alpha$ agrees perfectly with that from  H$\alpha$+22\,$\micron$ \citep{KE12}.

\begin{figure}
\includegraphics[trim=0mm 0mm 0mm 0mm,clip,width=0.47\textwidth]{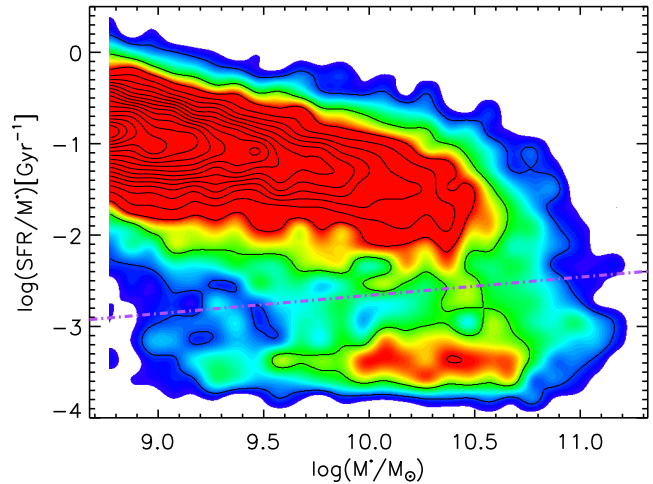}
\caption{Distribution of 21\,307 sample galaxies with $0.01<z<0.03$ in $M^\ast-$SSFR diagram. 17\,807 objects above the purple dot-dashed line are classified as SFGs.
\label{fig1}}
\end{figure}

Figure~\ref{fig1} shows the distribution of all 21\,307 sample galaxies in the $M^\ast-$SSFR diagram.  Quiescent galaxies are clearly separated from the main sequence of SFGs.
We split sample galaxies into five mass bins over 10$^{9.3} -10^{10.8}\,M_\odot$ and select SSFR values in each bin where the number density reaches minimum as separation points between the populations.
The best-fit line to the 5 separation points is
shown in Figure \ref{fig1}. Galaxies above the line are classified as SFGs, yielding 17\,807 SFGs to form an SFMS. Of them, 8036 are with $\log(M^\ast/M_\odot)>$9.5, of which 6\,004 ($75\%$) have 22\,$\micron$ detection. Below this mass cut we adopt SFR based on H$\alpha$ for all SFGs because 22\,$\micron$ detection rate rapidly declines.
We fit the SFMS and obtain the best fit as $\log(SFR) = (0.56\pm 0.02) \times \log(M^\ast/10^{10}\,M_\odot) - (0.47\pm0.01)$.

It is necessary to decompose the central structure from the star-forming disk of a galaxy when we consider a ``realistic'' SSFR, since classical bulges are generally red and dead \citep[e.g.][]{Drory07}.
Following \citet{Bluck14}, each galaxy is assumed to consist of a classical bulge ($n=4$) and an exponential disk ($n=1$).
Using the catalog of bulge+disk decompositions in $g$ and $r$ from \citet{Simard11} together with the mean color-$M^*$ relation from the VAGCs, we convert $r$-band bulge-to-total light ratio ($(B/T)_r$) of a galaxy into bulge-to-total mass ratio ($B/T$) and estimate stellar mass of the disk component ($M_D^*$) from $M^*$.  Our estimates of $B/T$ are consistent with those based on resolved broadband SEDs \citep{Mendel14}.
A large fraction of SFGs with $\log(M^\ast/M_\odot)>10.3$ in our sample are not included in
the bulge+disk decomposition catalog
because of $r>$14.  We use global S{\'{e}}rsic index to estimate $B/T$. Galaxies with global S{\'{e}}rsic index $n\le 1.5$ are treated as disk galaxies with $B/T$=0 and those with $n$ $\ge$ 4 are usually elliptical galaxies with $B/T$=1.  The mean relation between $B/T$ and $n$ from the bulge+disk decomposition catalog is used to estimate $B/T$ for galaxies with $n$ between 1.5 and 4.  Our sample SFGs with $\log(M^\ast/M_\odot)<$9.5 are mostly disk-dominated with $n<$1.5. By doing so, we obtain disk stellar mass $M_D^*$ for our sample SFGs.  Considering that star formation is in general irrelevant with the classical bulge of a galaxy, we divide total SFR by $M_D^*$ to obtain disk SSFR.

\begin{figure}
\includegraphics[trim=0mm 0mm 0mm 0mm,clip,width=0.47\textwidth]{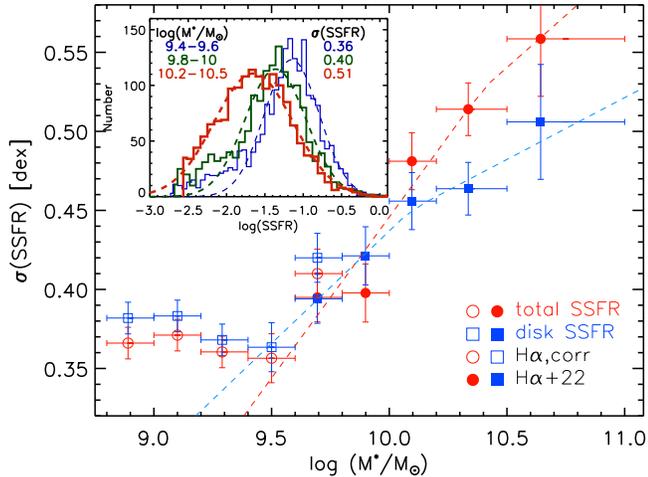}
\caption{Dispersion of SSFR as a function of total $M^*$ for SFGs.  Total SSFR (SFR divided by $M^*$) are marked by {\it red} circles, and disk SSFR (SFR divided by $M_D^*$) are shown in {\it blue} squares.
The solid and open symbols represent SFRs estimated from H$\alpha$+22\,$\micron$ and extinction-corrected H$\alpha$, respectively. 
\label{fig2}
}
\end{figure}

\section{The Dispersion of the Star-Formation Main Sequence}\label{sec:sfr}
Our 17807 sample SFGs form an SFMS at $z\sim 0.02$, best described by
\begin{equation}\label{eq:5}
 \log(SSFR_{\rm t}) = \alpha\times \log(M^\ast/10^{10.5}\,M_\odot) + \beta
\end{equation}
with $\alpha=-0.44$ and $\beta=-1.68$. Here $\rm SSFR_{\rm t}$ is total SSFR given in units of Gyr$^{-1}$. If disk SSFR
is used, then we obtain the best-fit SFMS with $\alpha=-0.32$ and $\beta=-1.42$.
For the relation between disk SSFR and $M_D^*$, the best-fit parameters are $\alpha=-0.35$ and $\beta=-1.54$.
It is clear that the relation between SSFR and $M^\ast$ becomes flattening when disk SSFR is adopted. Still, the disks in more massive SFGs have on average a lower SSFR and thus a lower level of star formation activity in general.

We split SFGs of 10$^{8.8-10.2}\,M_\odot$ into seven mass bins of width=0.2\,dex, and two bins $10^{10.2-10.5}\,M_\odot$ and $10^{10.5-11}\,M_\odot$. Fitting a normal distribution to the histogram of SSFR in logarithm for each mass bin, we take 1\,$\sigma$ of the best-fit profile as SSFR dispersion ($\sigma_{\rm SSFR}$) in units of dex.  Figure~\ref{fig2} shows the SSFR dispersion as a function of $M^*$.  It can be seen that the dispersion $\sigma_{\rm SSFR}$ increases with $M^*$ from $0.36\pm0.01$\,dex at $<10^{9.6}\,M_\odot$ to $0.51\pm0.02$\,dex at $>10^{10.2}\,M_\odot$.
 In other words, more massive SFGs have wider spread in SSFR. Similarly, the dispersion of disk SSFR is also measured. As shown in Figure~\ref{fig2}, the same increasing tendency is found for the dispersion of disk SSFR at $\log(M^\ast/M_\odot)>$9.6.
 At $\log(M^\ast/M_\odot)>$10.2,  the dispersion of disk SSFR is indeed lower than that of total SSFR although uncertainties are large.
It worth noting that the average disk SSFR is higher than the average total SSFR  at  $\log(M^\ast/M_\odot)>$10.2  but the dispersion of disk SSFR is smaller than that of total SSFR for the same sub-population of SFGs.  Accounting for measurement errors in $B/T$, the intrinsic dispersion in disk SSFR would be further smaller merely at the high-mass end.

The best-fit log-normal profiles of SSFR distribution in three different mass bins are shown in the inner panel of Figure~\ref{fig2}. The SSFR dispersion of SFGs with $\log(M^\ast/M_\odot)$=8.8$-$9.6 is based on extinction-corrected H$\alpha$. 
We notice that the dispersion of H$\alpha$-based SSFR is slightly larger than that based on H$\alpha+22\,\micron$. This discrepancy is likely due to the uncertainties in fiber aperture correction ($\sim$0.2\,dex) while WISE 22\,$\micron$ observation is free from such uncertainties.
Nevertheless, the dispersion of SSFR at the low-mass end with $\log(M^\ast/M_\odot)<$9.6 remains substantially lower than that at $\log(M^\ast/M_\odot)>$10.2.
We note that for the mass bin $10^{10.5-11}\,M_\odot$, SSFR dispersion is somewhat broadened by the systematic change in SSFR across the mass range, besides the increase of SFMS scatter itself. The latter stays $\sim 0.51$\,dex for $M^*>10^{10.2}M_\odot$.
Our estimate of SFMS scatter appears high compared to those in previous works, likely due to the systematic uncertainties in fiber effect correction, and sample selection as the wide coverage in the IR (and SFR) enables a proper selection of SFGs in terms of SSFR.

\begin{figure}
\includegraphics[trim=0mm 0mm 0mm 0mm,clip,width=0.47\textwidth]{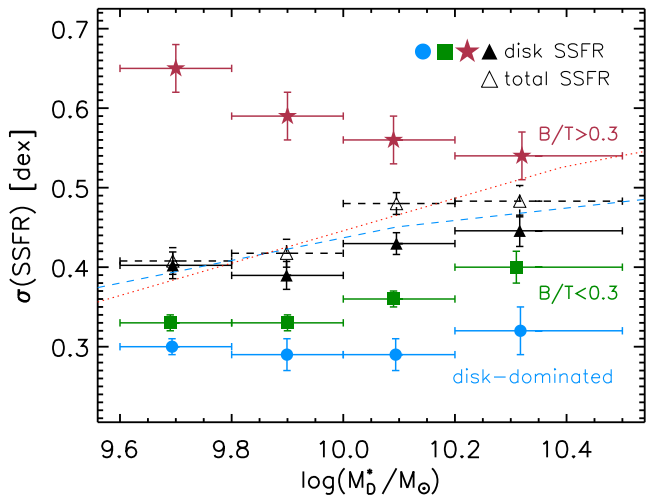}
\caption{
Dispersion of SSFR  as a function of $M_D^*$ for mass-selected SFGs ({\it triangles}) and the sub-samples of disk-dominated SFGs ({\it circles}), and SFGs with $B/T>0.3$ ({\it stars}) and with $B/T<0.3$ ({\it squares}), respectively.  {\it Open triangles} represent the dispersion in total SSFR, while the rest symbols refer to the dispersion in disk SSFR.  The dashed (dotted) line represents the best fit to the relation between the total (disk) SSFR dispersion and $M^*$ in Figure~\ref{fig2}.  \label{fig3}
}
\end{figure}

We further examine SSFR dispersion as a function of disk mass $M_D^*$. For comparison, we select a sub-sample of disk-dominated SFGs from our sample of SFGs using the criteria $(B/T)_{\rm r} \le 0.2$ and $n\le 1.5$.  
Similarly, subsamples of SFGs with $B/T>0.3$ and $B/T<0.3$ are drawn.  For the disk-dominated SFGs, $M^*$ equals to $M_D^*$. We split our SFG sample and the subsamples into the same nine mass bins in terms of $M_D^*$ and measure SSFR dispersion.
Figure~\ref{fig3} presents SSFR dispersion as a function of $M_D^*$ for SFGs, disk-dominated SFGs, SFGs with $B/T>0.3$ and with $B/T<0.3$, respectively. For the parent sample, dispersions in total SSFR and in disk SSFR are both calculated and shown. The comparison is limited to the four mass bins over $10^{9.6-10.5}\,M_\odot$, where SFR is robustly estimated in a consistent way. The dashed (dotted) line represents the best-fit relation between the total (disk) SSFR dispersion and $M^*$ in Figure~\ref{fig2}.
For mass-selected SFGs grouped by $M_D^*$ (triangles), the dispersion of total SSFR is similar to that obtained in bins split by $M^*$; and the dispersion of disk SSFR is noticeably smaller than the former in disks with $M_D^*>10^{10}\,M_\odot$.
Surprisingly, SSFR dispersion becomes substantially smaller and remains nearly constant (0.30\,dex) over the mass range $10^{9.5-10.5}\,M_\odot$ when only disk-dominated SFGs are counted.  When splitting the parent sample of SFGs by $B/T$, disk SSFR dispersion shows a distinct separation between the two sub-samples.  We point out that the cut $B/T>0.3$ selects the subsample of SFGs with $B/T$ and $M^*$ spreading over wide ranges; a smaller $M_D^*$ covers a wider range of $M^*$ with thus a larger dispersion in disk SSFR.  Meanwhile, SFGs with $B/T<0.3$ are close to disk-dominated SFGs with similar $M^*$ and less affected by central bulges. We conclude that the mixture of the disks in SFGs with different $B/T$ and $M^*$ maximizes disk SSFR dispersion, while the dispersion is minimized when only pure disks are counted.

\vspace{6pt}

\section{Discussion and Conclusion} \label{sec:discussion}

Our results show that disk SSFR is generally higher than total SSFR when global star formation is linked to the disk component of SFGs. Differing from the finding of mass independence for disk SSFR by \citet{Abramson14}, we obtain the $M^\ast-$ disk SSFR relation with a slope of $-0.32$, suggesting that disks in more massive SFGs have on average a lower level of star formation activity. A slope of $-0.35$ is obtained for the relation between disk SSFR and $M^\ast_D$.  Quenching of star formation in SFGs involves processes of consuming gas within galaxies and shutting down gas accretion in halos. The latter is strongly dependent on halo mass \citep{Woo13}, and may associate with slow quenching processes \citep{Fang13,Peng15}. In such a framework,
it is not surprising that disks in massive galaxies lack gas supply compared to low-mass SFGs , especially disk-dominated ones. On the other hand, massive SFGs tend to have a prominent central bulge \citep{Bluck14}, which may stabilize gas in disks and suppress star formation \citep[so-called morphology quenching,][]{Martig09}. We therefore argue that the decline of star formation in disks of massive SFGs is a natural consequence of halo quenching and probably affected by central bulges through AGN feedback or morphology quenching.

The most striking result  from Figure~\ref{fig2} is that SSFR dispersion increases remarkably with $M^*$,   from $\sigma_{\rm SSFR}=0.37\pm0.01$\,dex at $<10^{9.6}\,M_\odot$ to $0.51\pm0.02$\,dex at $>10^{10.2}\,M_\odot$. An increase with $M^*$ is also found for disk SSFR dispersion. The mass dependence of SSFR dispersion has been seen in SFGs at $z\sim 0.7$ \citep{Guo13} and over a wide redshift range \citep{Ilbert14}.
Dispersion in SSFR traces SFH diversity and variation of SFRs in SFGs of similar $M^*$. Here the SFH diversity refers to SFR offset over time scales of $>\sim 10^{8-9}$\,yr and the variation in H$\alpha$-based SFR is on time scales of  $\sim 10^7$\,yr.
We argue that the increase of SSFR dispersion is governed by mixture of physical processes regulating stochastic starbursts on short time scales and staged star formation on relatively-long time scales.  For massive SFGs,  halo-driven processes are proposed to suppress gas accretion and drive massive SFGs leaving SFMS slowly \citep{Peng15}; secular processes (disk instabilities, bar-driven tidal disruption, minor mergers) and major mergers/interactions induce starbursts followed by strong stellar feedback afterward; central bulges influence star formation in SFGs via bulge-driven processes
and the scatter in $B/T$ for SFGs with similar $M^*$ contribute additional spread to the SSFR dispersion. In contrast, lower-mass SFGs are less affected by these processes and exhibit a smaller SSFR dispersion in general.  In addition,  low-mass satellite galaxies are more sensitive to environmental quenching \citep{Peng10}. Since our sample is selected from a limited volume, the environmental effects on the SSFR dispersion are difficult to examine, leaving an open issue for future studies.

However, SSFR estimate based on mid-IR emission might be significantly contributed by old stellar population in massive galaxies \citep{Salim09,Chang15}, diluting the significance of the increment of SSFR dispersion with $M^*$ we found.
 Nevertheless, the dispersion of $H\alpha$-based SSFR is also confirmed to increase with $M^*$, having 0.55, 0.62, 0.65\,dex in the three mass bins over $10^{10-11}\,M_\odot$, respectively. These are much higher than the dispersions of $H\alpha+22\,\micron$-based SSFR. The reasons for this discrepancy are unclear. We suspect that complex dust geometry could be one reason.

A remarkable discrepancy in disk SSFR dispersion is found between the disk component of massive SFGs and morphology-limited disks (see Figure~\ref{fig3}), although measurement errors in $B/T$ may slightly enlarge the dispersion for the former. The comparison of dispersion in disk SSFR between $B/T>0.3$ and $B/T<0.3$ reveals that a mixture of SFGs with $B/T$ and $M^*$ over wider ranges leads to a larger spread in disk SSFR.  These again support that the existence of central bulges/bars enlarges the dispersion in disk SSFR. Interestingly, disk SSFR dispersion becomes substantially smaller and remains roughly constant over 9.6$<\log(M^\ast/M_\odot)<$10.5 when only disk-dominated SFGs are counted. The similarity of SSFR dispersion across $M_D^*$ for disk-limited SFGs is also seen at $z\sim$2 \citep{Salmi12}. This finding of mass-independence of SSFR dispersion strongly suggests that star formation in pure disks is free from bulge-driven processes, following the same pattern of star formation at least over 9.6$<\log(M^\ast/M_\odot)<$10.5.  The similarity also implies that the mode of stochastic and bursty star formation in disk-dominated SFGs is unlikely correlated with halo potential.
Our finding does not support the theoretical predictions in \citet{Hopkins14} that SSFR dispersion rapidly increases with decreasing $M^\ast$ as stellar feedback
plays an important role in driving a higher fraction of gas out of lower-mass galaxies and prevents further star formation.  This conflict could be caused by the treatment of stellar feedback in simulations if a fixed fraction of total energy output is set to heat interstellar medium (ISM). Stellar superwinds are often seen to move along the direction perpendicular to the disk plane of a starburst galaxy (e.g., M82) and might not strongly effect on ISM.
The mass-independence of SSFR dispersion in pure disks should be expected if the impact of stellar feedback on further star formation is negligible.

Our results suggest that the presence of central dense structures in massive SFGs has dramatic effects on galaxy-scale star formation, leaving not only the mean total/disk SSFR systematically lower but also the SSFR dispersion larger, compared to low-mass disk-dominated SFGs.  Our finding of the mass-independence of SSFR dispersion in pure disks indicates that disk galaxies of different $M^*$ obey the same mode of star formation, showing no correlation with halo potential.

\acknowledgments
We thank Yu-Yen CHANG for useful discussion. We are grateful to the anonymous referee for providing us with constructive comments and suggestions. This work is supported by the National Basic Research Program of China (973 Program 2013CB834900) and the Strategic Priority Research Program "The Emergence of Cosmological Structures" of the Chinese Academy of Sciences (Grant No. XDB09000000).

\clearpage

\end{document}